\documentclass[aps,prd,12pt,amssymb,amsmath,groupedaddress]{revtex4}
\usepackage{graphics}
\usepackage{url}

\begin{document}

\title{Observational Constraints on Theories with a Blue 
Spectrum of Tensor Modes}

\author{Andrew Stewart \footnote{stewarta@physics.mcgill.ca} and
Robert Brandenberger \footnote{rhb@physics.mcgill.ca}}
\affiliation{Physics Department, McGill University, Montreal, QC, H3A 2T8,
CANADA}

\date{\today}

\begin{abstract}
Motivated by the string gas cosmological model, which predicts a blue tilt of 
the primordial gravitational wave spectrum, we examine the constraints imposed 
by current and planned observations on a blue tilted tensor spectrum. Starting 
from an expression for the primordial gravitational wave spectrum normalised 
using cosmic microwave background observations, pulsar timing, direct 
detection and nucleosynthesis bounds are examined. If we assume a
tensor to scalar ratio on scales of the CMB which equals the current
observational upper bound, we obtain from these current observations
constraints on the tensor spectral index of $n_{T} \lesssim 0.79$, 
$n_{T} \lesssim 0.53$, and $n_{T} \lesssim 0.15$ respectively.
\end{abstract}

\maketitle

\section{Introduction}

A stochastic background of primordial gravitational waves represents valuable 
information about the very early universe as well as a way to discriminate 
between the myriad of cosmological models currently proposed. Although a 
stochastic background of gravitational waves has yet to be directly detected, 
efforts are being undertaken to do so by many current and future experiments,
some of the largest being LIGO, GEO600, TAMA, and LISA \cite{Allen:1996vm}. 
A stochastic background of gravitational waves is characterised by the 
gravitational wave spectrum
\begin{equation}
\Omega_{gw} \, = \, \frac{1}{\rho_{c}}\frac{d\rho_{gw}}{d\ln f}\,,
\end{equation}
where $\rho_{c}$ is the critical density of the universe, $\rho_{gw}$ is the 
energy density of the background gravitational waves and $f$ is 
frequency \cite{Allen:1996vm}. In the above, the energy density
in gravitational waves is written as an integral over $\ln f$, and the
derivative picks out the integrand.
In practice, the gravitational wave spectrum 
is commonly assumed to depend on frequency as a power of $f$. 

Within the framework of the inflationary universe scenario the primordial 
gravitational wave spectrum is predicted to be nearly scale invariant with a 
slight red tilt \cite{Spergel:2006hy}, i.e. more power at large scales. The
reason for the red tilt is that the amplitude of the gravitational wave
spectrum on a fixed scale $k$ is set by the Hubble constant $H$ at the time 
$t_i(k)$ when the scale $k$ exits the Hubble radius during the period
of inflation. Smaller scales exit the Hubble radius later when the Hubble
constant is smaller, leading to the red tilt. However, there do exist 
alternative cosmological models to the standard inflationary scenario which 
predict a blue tilt of the gravitational wave spectrum, i.e. more power at 
small scales, which have not yet been ruled out by observations. One 
cosmological model that predicts such a tilt is string gas cosmology \cite{BV}.
 
String gas cosmology \cite{BV} 
is an approach to string cosmology which starts from the new degrees of 
freedom and symmetries which string theory contains, but particle physics-based
models lack, namely string winding modes, string oscillatory modes and 
T-duality symmetry, and uses them to develop a new cosmological model. 
The claim is that by making these crucial stringy additions, one
obtains a new cosmological model which is singularity free \cite{BV}, 
generates nearly scale-invariant scalar metric perturbations from initial
string thermodynamic fluctuations \cite{NBV,BNPV1} and provides a natural 
explanation for the observed dimensionality of space \cite{BV} (see
\cite{RHBrev} for an overview of the string gas cosmology structure
formation scenario). A key
result which emerges \cite{BNPV2} from string gas cosmology 
is that the gravitational wave spectrum has a slight blue tilt, 
giving rise to a testable prediction different from that of the 
inflationary universe paradigm.

Using string gas cosmology as a motivation, in this work we will focus on 
how a tensor spectrum with a blue tilt can be constrained by current 
observational results, and what the prospects for improved constraints 
from some planned experiments are.

The starting point of our analysis will be an expression for the primordial 
gravitational wave spectrum normalised by cosmic microwave background (CMB) 
observations. Chongchitnan and Efstathiou \cite{Chongchitnan:2006pe} have 
derived just such an expression with a pivot scale 
$k_{0}=0.002\:\mbox{Mpc}^{-1}$ using the combined results from multiple 
surveys. Using the value $\mathcal{P}_{S}(k_{0})\simeq2.21\times10^{-9}$ 
for the amplitude of the scalar power spectrum evaluated at the pivot scale, they found $\Omega_{gw}(f)$ can be written as
\begin{equation}\label{chong}
h^{2}\Omega_{gw}(f) \, \simeq \, 
4.36\times10^{-15}r\left(\frac{f}{f_{0}}\right)^{n_{T}}\,,
\end{equation}
where $f_{0}=3.10\times10^{-18}$ Hz. Solving this result for the tilt of the gravitational wave spectrum, $n_{T}$, we get the explicit expression 
\begin{equation}\label{tilt}
n_{T} \, \simeq \, \frac{1}{\ln(f)-\ln(f_{0})}\ln\left(2.29\times10^{14}\frac{h^2\Omega_{gw}(f)}{r}\right)\,.
\end{equation}
In the above equations $r$ is the tensor-to-scalar ratio evaluated at the 
pivot scale,
\begin{equation}
r \, = \, \frac{\mathcal{P}_{T}(k_{0})}{\mathcal{P}_{S}(k_{0})}\,.
\end{equation}

The calculation of the tensor-to-scalar ratio depends quite sensitively on 
the parameters of the cosmological model under consideration. For that reason 
we choose to leave $r$ as a free parameter in the main expressions 
calculated in this work. However, for the sake of examining some numerical 
values of the constraints derived here we will insert a value of $r$ corresponding to the current upper bound into our results. Note, however,
that the bounds we derive depend only logarithmically on $r$. In each case we 
choose to use the value of the tensor-to-scalar ratio given by the combined three-year Wilkinson Microwave Anisotropy Probe (WMAP) and lensing normalised 
Sloan Digital Sky Survey (SDSS) data applied to a typical 
$\Lambda\mbox{CDM}$ model including tensors \cite{Spergel:2006hy}.  

In Sections II and III we will use the expression for the tilt of the 
gravitational wave
spectrum \eqref{tilt} along with several direct detection constraints 
to calculate the bounds on the required blue tilt of the spectrum. In 
Section IV we calculate the bound on the required blue tilt of the 
spectrum arising indirectly from the theory of big bang nucleosynthesis. In 
Section V we will investigate whether the CMB observations are compatible with
the calculated bounds, or if they will offer even tighter constraints. We 
conclude with a discussion of our findings along with other issues related 
to the method used.

\section{Pulsar Timing}

High precision measurements of millisecond pulsars provide a natural way to 
study low frequency gravitational waves. A gravitational wave passing between 
the earth and the pulsar will cause a slight change in the time of arrival of 
the pulse leading to a detectable signal.

\subsection{Parkes Pulsar Timing Array}

The Parkes Pulsar Timing Array (PPTA) project \cite{Manchester:2006xj} is 
a pulsar timing experiment using the Parkes 64m radio telescope located in
Australia with the ultimate goal of reaching the required sensitivity to make a
direct detection of gravitational waves. The PPTA project hopes to make 
timing observations of a sample of twenty millisecond pulsars, ten or more of 
which with a precision of less than approximately $100\:\mbox{ns}$. 

Jenet et al. \cite{Jenet:2005pv} have developed a technique to make a 
definitive
detection of a stochastic gravitational wave background by looking for 
correlations between pulsar observations. They have applied their method to 
data from seven pulsars observed by the PPTA project combined with an 
earlier data set to find a constraint on the amplitude of the characteristic
strain spectrum. They then used this result to place a bound on the 
primordial gravitational wave spectrum \cite{Jenet:2006sv}
\begin{equation}
h^{2}\Omega_{gw}(1/8\:\mbox{yr}) \, \leq \, 2.0\times10^{-8} \, .
\end{equation}

Plugging the constraint of Jenet et al. into Equation \eqref{tilt} at 
the frequency 
$f=1/8\:\mbox{yr}\simeq3.96\times10^{-9}\:\mbox{Hz}$ we obtain a constraint
\begin{equation}
n_{T} \, \lesssim \, 0.0477\ln\left(\frac{4.59\times10^{6}}{r}\right) \, .
\end{equation}
The WMAP+SDSS data places a bound $r<0.30$ on the tensor-to-scalar ratio 
\cite{Spergel:2006hy}. Inserting $r=0.30$ into the above equation we find 
that the current pulsar timing observations constrain the blue tilt of the 
tensor spectrum to $n_{T} \lesssim 0.79$. 

Jenet et al. have also used simulated data to determine the upper bound on 
the primordial gravitational wave spectrum expected from future pulsar observations.
Using a simulated data-set of twenty pulsars timed with an rms timing residual 
of $100\:\mbox{ns}$ over 5 years they calculated \cite{Jenet:2006sv}
\begin{equation}
h^{2}\Omega_{gw}(1/8\:\mbox{yr}) \, \leq \, 9.1\times10^{-11} \, .
\end{equation}
Plugging this improved constraint into Equation \eqref{tilt} at the frequency $f=1/8\:\mbox{yr}$ we get a bound
\begin{equation}
n_{T} \, \lesssim \, 0.0477\ln\left(\frac{2.09\times10^{4}}{r}\right) \, ,
\end{equation}
and again using the value $r=0.30$, we find that, in the absence of a detection,
future pulsar timing observations could tighten the constraint on the blue 
tilt to $n_{T} \lesssim 0.53$.

\section{Interferometers}

Interferometer experiments offer a way to directly measure the gravitational 
wave strain spectrum with many observatories currently running or planned for 
the future. Interferometers in different locations form a network that 
will search for a correlated signal between detectors beneath 
uncorrelated detector noise, in order to improve sensitivity.

\subsection{LIGO}

The Laser Interferometer Gravitational Wave Observatory (LIGO) 
\cite{Abbott:2006zx} is a ground based interferometer project operating 
in the frequency range of $10\:\mbox{Hz}$ - a few kHz. LIGO consists of 
two collocated Michelson interferometers in Hanford, Washington, H1 with 
$4\:\mbox{km}$ long arms, and H2 with $2\:\mbox{km}$ long arms, along with 
a third interferometer in Livingston Parish, Louisiana, L1 with 
$4\:\mbox{km}$ long arms. 

Most recently LIGO has performed its fourth 
science run, S4, with improved interferometer sensitivity. Abbott et al. 
\cite{Abbott:2006zx} have used the S4 data to calculate a limit on 
the amplitude of a frequency independent gravitational wave spectrum. 
They found a bound 
\begin{equation}
\Omega_{gw} \, < \, 6.5\times10^{-5}
\end{equation} 
in the frequency range 51-150 Hz. Inserting this value into 
Equation \eqref{tilt} at the frequency $f=100\:\mbox{Hz}$ we get a 
constraint
\begin{equation}\label{r2}
n_{T} \, \lesssim \, 0.0223\ln\left(\frac{1.49\times10^{10}h^{2}}{r}\right)\,.
\end{equation}
The WMAP+SDSS data also provides a value of $h=0.716$ for the Hubble 
parameter \cite{LAMBDA}. Inserting this along with $r=0.30$ into the above bound, 
we find that the current LIGO results place a constraint $n_{T}\lesssim0.53$ 
on the blue tilt of the tensor spectrum. 

The final phase of LIGO, named Advanced LIGO, hopes to reach a detection 
sensitivity of \cite{Abbott:2006zx}
\begin{equation}
\Omega_{gw} \, \sim \, 10^{-9} \, . 
\end{equation}
Plugging this value into Equation \eqref{tilt} at $f=100\:\mbox{Hz}$, we 
find that if Advanced LIGO does not make a positive detection of a gravitational wave background then it will place a bound on the 
blue tilt of the tensor spectrum
\begin{equation}
n_{T} \, \lesssim \, 0.0223\ln\left(\frac{2.29\times10^{5}h^{2}}{r}\right)\,.
\end{equation}
Using $r=0.30$ and $h=0.716$ in this expression, Advanced LIGO
would then constrain the blue tilt of the tensor spectrum to 
$n_{T} \lesssim 0.29$.

\subsection{LISA}

The Laser Interferometer Space Antenna (LISA) \cite{Danzmann:2003tv} is a 
planned space-based interferometer experiment operating in the mHz range. 
LISA will consist of three drag-free spacecraft each at the corner of 
an equilateral triangle with sides of length $5\times10^{9}\:\mbox{m}$. 
Each spacecraft has two optical assemblies pointed towards the other two
spacecraft forming three Michelson interferometers. This triangle formation 
will orbit the sun in an Earth-like orbit separated from us by approximately
fifty million kilometers. The goal of LISA is to reach a sensitivity of
\cite{Maggiore:2000gv}
\begin{equation}
h^{2}\Omega_{gw}(1\:\mbox{mHz}) \, \simeq \, 1 \times10^{-12} \, . 
\end{equation}
At the LISA sensitivity level one would expect gravitational wave signals 
from super-massive black hole binaries, other binary systems and 
super-massive black hole formation to be present. Assuming these 
predicted signals could somehow be removed and LISA does not detect 
any primordial signal, we can plug this predicted limit into 
Equation \eqref{tilt} at $f=1\:\mbox{mHz}$ to obtain 
a limit on the blue tilt of the primordial gravitational wave spectrum
\begin{equation}
n_{T} \, \lesssim \, 0.0299\ln\left(\frac{2.29\times10^{2}}{r}\right)\,.
\end{equation}
Inserting $r=0.30$ into this equation we find that LISA could potentially 
place a constraint $n_{T}\lesssim0.20$ on the blue tilt.

\section{Nucleosynthesis}

The theory of big-bang nucleosynthesis (BBN) successfully predicts the 
observed abundances of several light elements in the universe. In doing so, 
BBN places constraints on a number of cosmological parameters. This in 
turn results in an indirect constraint on the energy density in a 
gravitational wave background as follows: the presence of a significant 
amount of gravitational radiation at the time of nucleosynthesis will 
change the total energy density of the universe, which affects the rate 
of expansion in that era, leading to an over-abundance of helium and 
thus spoiling the predictions of BBN \cite{Allen:1996vm}. Assuming 
$N_{\nu}=4.4$, where $N_{\nu}$ is the effective number of neutrino species 
at the time of nucleosynthesis, the BBN bound is \cite{Abbott:2006zx}
\begin{equation}
\int^{f_{2}}_{f_{1}}\Omega_{gw}(f)\,d(\ln f) \, < \, 1.5\times10^{-5}\,.
\end{equation}
Plugging Equation \eqref{chong} into the left-hand side and performing 
the integration we obtain the inequality
\begin{equation}\label{BBN}
\frac{f^{n_{T}}_{2}-f^{n_{T}}_{1}}{n_{T}} \, < \,
3.4\times10^{9}\,\frac{h^{2}f^{n_{T}}_{0}}{r}\,.
\end{equation}

In order to apply the above result, we must discuss the two integration
limits $f_1$ and $f_2$. The lower cutoff frequency $f_1$ corresponds to
the Hubble radius at the time of BBN and takes the value 
$f_{1}\sim10^{-10}\:\mbox{Hz}$. For wavelengths larger than the
Hubble radius, the gravitational waves are frozen out \cite{Grishchuk}
(see e.g. \cite{MFB} for a review) and thus do not act like radiation.
The upper cutoff frequency $f_2$ is the ultraviolet cutoff. We will 
take it to be given by the Planck frequency, i.e. 
$f_{2}=f_{Pl}=1.86\times10^{43}\:\mbox{Hz}$. Substituting these two limits 
into Equation \eqref{BBN} along with the WMAP+SDSS values $r=0.30$ and 
$h=0.716$ then solving numerically for $n_{T}$, we find the bound on the 
blue tilt of the tensor spectrum from BBN to be 
\begin{equation}
n_{T} \, \lesssim \, 0.15 \, .
\end{equation}
Had we instead inserted for $f_2$ the scale of grand unification, $10^{16}\:\mbox{GeV}$, or the Hubble rate during a simple large field inflation model, which is
$10^{13}\:\mbox{GeV}$, the bound would be slightly relaxed to
$0.16$ or $0.17$, respectively. Thus, the dependence of the bound
on the uncertain ultraviolet cutoff scale $f_{2}$ is quite mild.

Lastly, we do not want $\Omega_{gw} > 1$ at any scale within the integration bounds. Since we are working with such large frequencies we should check to be sure that this condition is satisfied. Substituting the value of the tensor spectral index determined by BBN back into Equation \eqref{chong} we find that $\Omega_{gw}(f_{Pl})=3.93\times10^{-6}$, meaning our requirement is indeed satisfied for all frequencies within the interval of integration.

\section{Cosmic Microwave Background}

Observations of the CMB have implications for a wide variety of topics 
including constraining inflation, dark matter and large-scale structure. 
The mission of WMAP is to produce full-sky maps of the CMB anisotropy and 
the recently published three-year results are an improvement upon previous 
observations. A reduction in instrument noise produced spectra which are three 
times more sensitive in the noise limited region, independent years of data 
allow for cross-checks, the instrument calibration and response have been 
better characterised and a thorough analysis of the polarisation data 
has improved the understanding of the data \cite{Hinshaw:2006ia}. Using three 
year WMAP data, the derived angular power spectrum of the temperature 
anisotropy, $C^{TT}_{l}$, where $l$ is the multipole moment, is cosmic 
variance limited to $l=400$ and the signal to noise ratio exceeds unity 
to $l=1000$ \cite{Hinshaw:2006ia}. This high precision cosmological data 
provides another method of placing constraints on the value of the tensor 
spectral index. 

Using the Code for Anisotropies in the Microwave Background (CAMB) 
\cite{Lewis:1999bs} we can simulate how a blue tilt of the primordial 
gravitational wave background would effect the anisotropies in the CMB. 
To examine possible constraints we employ the following method: first, 
we calculate $C^{TT}_{l}$ using CAMB for each of the three current bounds 
on $n_{T}$ calculated in the previous sections; second, we calculate 
$C^{TT}_{l}$ using CAMB, this time with a standard inflationary relation: 
$n_{T}=-r/8$ \cite{Spergel:2006hy}; finally, we compare the output 
$C^{TT}_{l}$ data for the models with a blue tilt against the output for 
the model with the ``usual'' value of the tensor spectral index (the
above relation from inflationary cosmology). For consistency with the 
previous sections, when running CAMB we choose our input cosmological 
parameters to be those calculated using the WMAP+SDSS data for a 
$\Lambda\mbox{CDM}$ model with tensors \cite{LAMBDA}.

\begin{figure}
\includegraphics{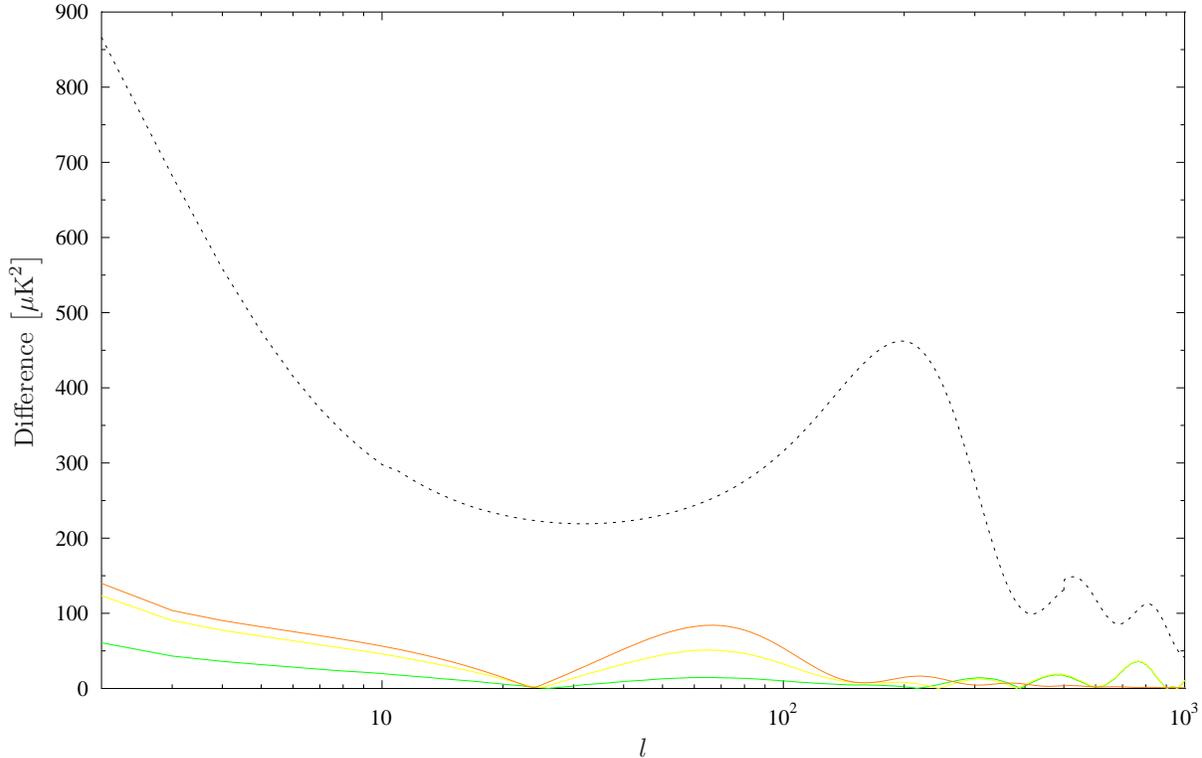}
\caption{\label{TT}Magnitude of the difference between the the 
TT power spectrum for a model with a blue tensor spectral index and the 
TT power spectrum for a model with a tensor spectral index defined as: 
$n_{T}=-r/8$. Shown here are the cases $n_{T}=0.15$ (green), $n_{T}=0.53$ 
(yellow) and $n_{T}=0.79$ (orange) respectively. The dashed line represents 
the cosmic variance error at each $l$.}
\end{figure}

{F}rom Figure \ref{TT} we can clearly see that the power spectrum of the 
temperature anisotropy for models with a blue tensor spectral index does 
not vary much from that calculated using a standard inflationary definition 
of the tensor spectral index. In fact, the difference is within the cosmic 
variance error at all $l\leq1000$ for each of the three bounds calculated 
using the PPTA observations, LIGO observations and the theory of BBN. 
Thus, we find that the cosmic microwave background does not offer any 
tighter constraints on the blue tilt of the gravitational wave spectrum 
than those already calculated.

\section{Conclusions}

A stochastic background of primordial gravitational waves is a prediction 
of many cosmological models. Assuming the gravitational wave spectrum 
depends as a power on frequency, then this spectrum can be
characterised by its tilt and amplitude. Most models predict the tilt 
to be nearly scale-invariant but slightly red, while some models, like string
gas cosmology, predict a slight blue tilt. Although a gravitational wave 
background has yet to be directly detected, observational results can 
already be used to constrain it. 

Using the current results from pulsar timing observations, direct detection observations and the theory of nucleosynthesis, we have placed bounds on the possible blue tilt of the gravitational wave background. After completion of this work we became aware of \cite{Boyle} in which a master equation was derived which relates the short wavelength observable $\Omega_{gw}(f)$ to the tensor to scalar ratio measured with the CMB. The goal of that work was to develop a formulation which is as general as possible. In particular, the equation of state parameter and the tensor spectral index are taken to be arbitrary functions of the scale factor and wavenumber respectively, not constants as is often assumed. Their master equation is thus a more general version of our ``master'' equation \eqref{chong} and we could have just as easily used it as our starting point. In fact, we have confirmed that by choosing a constant value $w=1/3$ for the equation of state parameter (which is described as the most logical value in \cite{Boyle}) and numerical values for other cosmological parameters that match our choices above, we can indeed re-derive all of our results from the master equation of \cite{Boyle}. We consider this a good consistency check for the constraints derived in this work. The authors of \cite{Boyle} do include a discussion of some of the same types of observations mentioned here, namely laser interferometer and pulsar timing, however, we stress that they do not use the current numerical constraint from any particular observatories to compute actual upper bounds on $n_T$, which was the purpose of this work. The authors of \cite{Boyle} also discuss a constraint on $n_T$ coming from BBN, but they take the constraint on $\Omega_{gw}$ from BBN to be a constant across all frequencies rather than integrating their master equation as was done in this work. In the end they find a weaker bound on the tilt, $n_T\lesssim0.36$, than the one obtained here.

By far the tightest constraint on the tilt comes from big-bang nucleosynthesis. If we take the tensor to scalar ratio on CMB scales to be given by the current observational upper bound, and if we take the ultraviolet cutoff scale in the spectrum of gravitational radiation to be the Planck scale, then the bound is $n_{T}\lesssim 0.15$, tighter than even Advanced LIGO, the future PPTA, and LISA can hope to achieve. It is not surprising that nucleosynthesis provides the tightest bounds on a blue tilt of the gravitational wave spectrum since nucleosynthesis probes physics on scales much smaller than the other experiments we analysed, and spectra with blue tilts have more power on the smallest scales. That is, BBN gives us the largest ``lever arm'' to probe gravitational waves in conjunction with CMB observations, a point also made in \cite{Peiris}. From our results we can clearly see the trend that the bound on the blue tilt of the tensor spectral index tightens as the length scale probed by the given experiment decreases.

Simulations of the angular power spectrum of the temperature anisotropies in the CMB did not offer any tighter constraints on the tensor spectral index, with each of the constraints calculated in this paper producing a temperature power spectrum that was within the cosmic variance error of one calculated for a standard $\Lambda\mbox{CDM}$ model with tensor modes included.

As mentioned in the Introduction, Equation \eqref{chong} has been normalised 
at the scale of cosmic microwave background observations. However, those 
experiments probe scales that are approximately ten orders of magnitude 
larger than those probed by the PPTA and approximately nineteen orders of 
magnitude larger than those probed by LIGO, with LISA probing between the two. 
Extrapolating between such a large difference in scales is not 
straightforward and we should note that in \cite{Chongchitnan:2006pe} the 
authors conclude from their analysis that even within the framework of the 
inflationary universe paradigm the formula for the primordial gravitational 
wave spectrum \eqref{chong} is too restrictive, and they believe it is 
indeed not possible to extrapolate reliably over such a large difference 
in scales. Whether or not this is the case in the string gas cosmology 
model should perhaps be examined more carefully in future work.

Continuing with string gas cosmology, we conclude that the current bounds
on the tilt of the gravitational wave spectrum are weak. The predicted
magnitude of the blue tilt of the gravity wave spectrum is thought
to be comparable to the magnitude of the red tilt of the spectrum of
scalar metric fluctuations \cite{BNPV2}. If the latter is taken to agree
with the current bounds, we predict a blue tilt of less than $n_T = 0.1$
which will not be easy to detect. There may, however, be models similar
to string gas cosmology in which the scalar and tensor tilts are 
not related, and for which planned experiments could set valuable
constraints on the model parameter space.

\section*{Acknowledgements}

This research is supported in part by NSERC Discovery Grant, by funds from the Canada Research Chairs program and by a FQRNT Team Grant. We acknowledge the use of the Legacy Archive for Microwave Background Data Analysis (LAMBDA). Support for LAMBDA is provided by the NASA Office of Space Science. We wish to thank Gil Holder for useful discussions.


\end{document}